\documentclass[a4paper,11pt]{article}
\usepackage{booktabs} 
\usepackage{graphicx} 
\usepackage{color}

\ifx\pdfminorversion\undefined\else\pdfminorversion=7\fi
\ifx\pdfsuppresswarningpagegroup\undefined\else\pdfsuppresswarningpagegroup=1\fi

\usepackage{pos,multirow,cancel,slashed,ucs,graphicx}
\graphicspath{{FIGS/}{Notes/}}
\colorlet{yellow}{yellow!70!black}

\usepackage[capitalize]{cleveref}

\makeatletter

\newcount\Hy@sublinkcounter
\newcommand\phantomsubsection{%
  \Hy@GlobalStepCount\Hy@sublinkcounter
  \xdef\@currentHref{subsection*.\the\Hy@linkcounter.\the\Hy@sublinkcounter}%
  \Hy@raisedlink{\hyper@anchorstart{\@currentHref}\hyper@anchorend}}
\expandafter\renewcommand\expandafter\phantomsection\expandafter
  {\phantomsection\global\Hy@sublinkcounter0\relax}

\@ifundefined{texorpdfstring}{\newcommand\texorpdfstring[2]{#1}}{}

\@ifundefined{unichar}
{\newcommand\myunichardef[3]{\expandafter\providecommand\csname text#1\endcsname
                            {#2}}}
{\newcommand\myunichardef[3]{\expandafter\providecommand\csname text#1\endcsname
                            {\unichar{"#3}}}}
\makeatother
\myunichardef{composetilde}{\string~}{0303}
\myunichardef{composemacron}{bar}{0305}
\myunichardef{subscriptthree}{\string_3}{2083}
\myunichardef{superscripttwo}{\string^2}{00B2}
\myunichardef{Theta}{Theta}{0398}

\usepackage[T1]{fontenc}

\usepackage{relsize}

\usepackage[acronyms, nohypertypes={acronym}, nopostdot, style=super, nonumberlist, toc]{glossaries}
\glsenableentrycount

\newcommand\ac[1]{\gls{#1}}

\newacronym{qcd}{QCD}{quantum chromodynamics}
\newacronym{bsm}{BSM}{beyond the standard model}
\newacronym{cp}{CP}{simultaneous interchange of left with right and particle with its antiparticle}
\newacronym{edm}{EDM}{electric dipole moment}
\newacronym{mdm}{EDM}{magnetic dipole moment}
\newacronym{cedm}{cEDM}{chromoelectric dipole moment}
\newacronym{cmdm}{cMDM}{chromomagnetic dipole moment}
\newacronym{qcedm}{qcEDM}{quark {\ac{cedm}}}
\newacronym{qcmdm}{qcMDM}{quark {\ac{cmdm}}}
\newacronym{qedm}{qEDM}{quark {\ac{edm}}}
\newacronym{nedm}{nEDM}{neutron {\ac{edm}}}
\newacronym{hisq}{HISQ}{highly improved staggered quark}
\newacronym{CPV}{CPV}{{\ac{cp}}-violation}
\newcommand\CPV{$\cancel{\text{CP}}$}
\newacronym{cpv}{\noexpand\CPV}{{\ac{cp}}-violating}
\newacronym{awi}{AWI}{axial Ward identity}
\newacronym{esc}{ESC}{excited state contamination}
\newacronym{chipt}{\(\chi\)PT}{chiral perturbation theory}
\newacronym{ckm}{CKM}{Cabbibo-Kobayashi-Maskawa quark-mixing}
\newacronym{mimd}{MIMD}{Multiple Instruction, Multiple Data}
\newacronym{milc}{MILC}{the Multiple Instruction, Multiple Data (MIMD) Lattice Computation}
\newacronym{ccfv}{CCFV}{chiral, continuum, and finite volume}
\newacronym[longplural=effective field theories]{eft}{EFT}{effective field theory}
\newacronym{pq}{PQ}{Peccei-Quinn}
\newacronym{aic}{AIC}{Akaike information criterion}

\title{Gradient flow of the Weinberg operator}

\author*[a]{Tanmoy Bhattacharya}
\author[b, a]{Shohini Bhattacharya}
\author[c]{Vincenzo Cirigliano}
\author[a]{Rajan Gupta}
\author[a]{Emanuele Mereghetti}
\author[d, e]{Sungwoo Park}
\author[a]{Jun-Sik Yoo}
\author[f]{Boram Yoon}

\affiliation[a]{Los Alamos National Laboratory,
MS B285, P.O. Box 1663, Los Alamos, NM 87545-0285, USA}

\affiliation[b]{Department of Physics, University of Connecticut, Storrs, CT 06269, USA}

\affiliation[c]{Physics Department, University of Washington,
  3910 15th Avenue NE, Seattle, WA 98195-1560, USA}

\affiliation[d]{Physical and Life Sciences Division, Lawrence Livermore National Lab, 7000 East Ave, Livermore, CA 94550, USA}
\affiliation[e]{Nuclear Science Division, Lawrence Berkeley National Laboratory, Berkeley, CA 94720, USA}

\affiliation[f]{NVIDIA Corporation, Santa Clara, CA 95050, USA}

\emailAdd{tanmoy@lanl.gov}
\emailAdd{shohinib@uconn.edu}
\emailAdd{cirigv@uw.edu}
\emailAdd{rg@lanl.gov}
\emailAdd{emereghetti@lanl.gov}
\emailAdd{park49@llnl.gov}
\emailAdd{junsik@lanl.gov}
\emailAdd{byoon@nvidia.com}

\abstract{We present preliminary results on the susceptibilities involving the CP-violating (CPV) Weinberg three-gluon operator and the topological $\Theta$ term using the gradient flow scheme, and study their continuum and chiral extrapolations. These are used to 
provide an estimate of the \(\Theta\) induced by the Weinberg operator in theories with the Peccei-Quinn (PQ) mechanism. Combined with the calculations of the matrix elements (MEs) of quark-bilinears between nucleon states, such calculations will enable estimates of the electric dipole moments (EDMs) and CPV pion-nucleon couplings due to the Weinberg operator, thereby providing robust constraints on beyond the standard model (BSM) physics.}

\FullConference{The 41st International Symposium on Lattice Field Theory (LATTICE2024)\\
 28 July - 3 August 2024\\
Liverpool, UK\\}

\renewcommand{\printHeadAuthors}{Bhattacharya et al.}

\begin{document}
\maketitle

\section{The case for physics Beyond the Standard Model}
Both the Standard Model (SM) of Particle Physics and the Standard Model of Cosmology have revolutionized our understanding of the universe, yet they are fundamentally incompatible in several critical areas, 
including the unexplained nature of dark energy, dark matter, and the origin of a matter-dominated universe.
According to the Standard Model of particle physics, 
the matter-antimatter asymmetry has to arise after the inflationary phase of the universe~\cite{Coppi:2004za}. 
It is well known that this requires violation of the CP symmetry (the combined symmetry of charge conjugation and parity) in the elementary interactions~\cite{Sakharov:1967dj}. While the Standard Model exhibits CP-violation through the Cabibbo-Kobayashi-Maskawa (CKM) quark-mixing matrix~\cite{Kobayashi:1973fv} in weak interactions---and through the similar leptonic Pontecorvo–Maki–Nakagawa–Sakata (PMNS) mixing matrix in the neutrino sector~\cite{Maki:1962mu,Nunokawa:2007qh}, the observed magnitude is probably too small to account for the observed  matter-antimatter asymmetry of the universe (BAU)~\cite{Shaposhnikov:1987tw,Farrar:1993sp,Gavela:1993ts,Gavela:1994dt,Gavela:1994ds,Huet:1994jb}. This deficiency  points to the existence of new sources of CP violation beyond the Standard Model (BSM), and finding experimental evidence for such theories is an important arena of current research.

\section{Electric Dipole Moments: Unambiguous signals of BSM physics}
Electric Dipole Moments (EDMs) are among the most sensitive tools for detecting BSM physics, particularly in the context of CP violation~\cite{Alarcon:2022ero}. Since nondegenerate quantum eigenstates do not have EDMs in CP-conserving theories, the detection of an EDM of an elementary particle would be a definite signal of CPV. Current experiments have put stringent limits on the EDMs of the electron~\cite{Roussy:2022cmp}, neutron~\cite{Abel:2020pzs}, and proton~\cite{Graner_2016} of $4.1 \times 10^{-30} \, e \cdot \text{cm}, 1.8 \times 10^{-26} \, e \cdot \text{cm}$ and $2.1 \times 10^{-25} \, e \cdot \text{cm}$ respectively, all at or within a few orders of magnitude of what would be na\"\i{}vely expected from a BSM CPV near the weak scale~\cite{Alarcon:2022ero}.  Importantly, CPV in the SM predicts EDMs to be extraordinarily small---roughly a million times weaker than the current experimental limits.\looseness-1

What makes EDMs particularly compelling today is the promise of significant near-term progress in their detection. Advances in experimental techniques are expected to improve sensitivity by 2 to 3 orders of magnitude~\cite{Alarcon:2022ero}. This increased precision has profound implications: it could allow us to either observe EDMs directly or rule out several leading BSM mechanisms for generating the CP violation necessary for baryogenesis.  

\section{Leading CPV operators}
An useful way of classifying the CPV operators is in terms of  their mass dimensions in the effective field theory (EFT) description of BSM physics~\cite{Engel:2013lsa}. Each operator opens a distinct window into CP-violation beyond the SM, with their importance shaped by energy scales, experimental reach, and theoretical context.

At the QCD scale of about a GeV,  low dimension operators are closely related to  SM physics. In particular, complex phases in the fermion mass matrix are of dimension 3 while the topological $\Theta$-term in QCD is of dimension 4.  These two are related: anomalous axial symmetry of the standard model rotates the topological term into the complex phases of the quark mass-matrix, implying their physical effects vanish if any quark is massless. Even with the observed small quark masses, an O(1) value of $\Theta$ would induce EDMs for the neutron a billion times its experimental constraint~\cite{Pospelov:2000bw}. Various models have been proposed to explain this, a leading contender being the Peccei-Quinn mechanism~\cite{Peccei:1977hh} that dynamically relaxes \(\Theta\) to zero in the absence of any other CPV interaction in the theory.  In contrast, the dimension-5 operators, suppressed by a high scale, arise from BSM physics. These are the EDMs for leptons and quarks, and the quark chromo-electric dipole moment. The effect of these on the nucleon EDMs are being  studied~\cite{Bhattacharya:2015esa,Gupta:2018lvp,Bhattacharya:2023qwf}.

These dimension-5 operators, in fact, arise from dimension-6 interactions involving the Higgs' field at the weak scale. While dimension-5, in many models they are suppressed by small Yukawa couplings~\cite{Pospelov:2000bw}. For this reason, their effects are often competitive with those that remain dimension-6 at the QCD scale: i.e., the CPV Weinberg three-gluon operator and various four-fermion interactions---some of which appear on integrating out the W and Z bosons in the standard model. The four-fermion operators are implicated in processes such as lepton-flavor violation, neutral meson mixing, and rare decays, but there is little work on the nonperturbative effects of any of these dimension 6 operators~\cite{Bhattacharya:2022whc}. Here we study the first of these: the CPV Weinberg operator that describes the chromoelectric moment of the gluon.

\section{Effective Field Theory, Renormalization, and Gradient Flow}
\subsection{Effective Field Theory: Separation of scales}
One of the reasons for the difficulty in understanding the effects of the higher-dimensional operators in effective field theory (EFT) is their nonperturbative dependence on the renormalization scheme. The EFT expansion of a typical ME in the BSM reads
\begin{align}
{\rm BSM\ ME} &= {\rm Wilson \, Coeff_1}\times{\rm ME(O_1)} +  \frac{\rm Wilson \, Coeff_2}{{\rm BSM\ scale}^n}\times{\rm ME(O_2)} + \ldots
\label{e:mix}
\end{align}
Here, $\rm O_1$ and $\rm O_2$ are operators in the EFT, with $\rm O_2$ having a mass dimension \(n\) higher than $\rm O_1$. Wilson coefficients (e.g., ${\rm Wilson \, Coeff_1}$ and ${\rm Wilson \, Coeff_2}$), which parametrize the short-distance effects, depend on the strong coupling constant $\alpha_s$ at the BSM scale. These coefficients are typically calculated using perturbation theory without a hard cutoff (e.g., in schemes using dimensional regularization), and the resulting expressions typically turn out to be non-Borel summable, leading to `ultraviolet renormalon' ambiguities in \({\rm Wilson \, Coeff_1}\) proportional to
\begin{equation}
   \exp \bigg (-\dfrac{n}{2\beta_0\alpha_s^2({\rm BSM\ scale})} \bigg ) \sim \bigg (\dfrac{\Lambda_{\rm QCD}}{{\rm BSM\ scale}} \bigg )^n \, ,
\end{equation}	
where $\Lambda_{\rm QCD}$ is the QCD scale, and $\beta_0$ is the first coefficient in the QCD beta function~\cite{Beneke:1998ui}. Consistent with this, if one attempted to compute \({\rm ME(O_2)}\) in perturbation theory without a hard cutoff, then one would find a canceling `infrared renormalon' proportional to \(({\Lambda_{\rm QCD}})^n\).  

A problem, however, arises if we attempt to merge the perturbative calculation of the Wilson coefficients with the nonperturbative evaluation of the MEs of the EFT operators, since the two pieces individually are ill-defined. In particular, any redefinition of the higher-dimensional operators of the form
\begin{equation}
   {\rm O_2} - {\rm MixingCoeff_{21}} \, \Lambda_{\rm QCD}^n\; {\rm O_1} \, .
\end{equation}	
can be absorbed into a nonperturbatively small change in the ${\rm Wilson \, Coeff_1}$. This is, however, no more than a scheme dependence: Any choice  that consistently calculates both the Wilson coefficients and defines the higher-dimensional operators in Eq.~(\ref{e:mix}) can be used to calculate the BSM MEs from the EFT operators with no ambiguity up to the given order in \(({\Lambda_{\rm QCD}}/{{\rm BSM\ scale}})\).

\subsection{Renormalization and power divergences in EFT}
As described in the previous subsection, higher-dimensional operators in EFT (e.g., $\rm O_2$ in Eq.~(\ref{e:mix})) obtain a non-perturbative scheme-dependent contribution proportional to the lower-dimensional operators. In particular, their MEs regularized with a hard cutoff often exhibit power divergences as the cutoff (the scale above which degrees of freedom are integrated out) increases. These divergences grow as ${\rm Cutoff}^n$, where $n$ depends on the operator's dimension, and, provided the action is properly renormalized, are proportional to the MEs of lower-dimensional operators, such as ${\rm ME}({\rm Cutoff}^n {\rm O_1})$. To remove the cutoff, one needs to renormalize the operators with a subtraction: \({\rm O_2} - {\rm MixingCoeff_{21}} {\rm Cutoff}^n {\rm O_1}\), and the scheme dependence arises since a change of \({\rm MixingCoeff_{21}}\) by a piece vanishing as \({\rm Cutoff}^{-n}\) changes the renormalized MEs by a finite amount.

\subsection{Gradient Flow scheme}
As discussed above, higher-dimensional operators in `hard cutoff' theories are power-divergent, and the subtraction of this power divergence introduces scheme dependence.  An especially convenient scheme for regulating this divergence is the `Gradient Flow renormalization scheme'~\cite{Luscher:2010iy}. In this scheme, one introduces a hard cutoff $\tau_{\rm gf} = \sqrt{8 t_{gf}} a$, where $a$ is the lattice spacing that sets the cutoff in lattice regularization. All fields are smeared in position space to this \(\tau_{\rm gf}\) scale.  In particular, the gauge field $U(t_{\rm gf})$ is evolved according to the gradient flow equation:
\begin{equation}
\frac{\partial U(t_{\rm gf})}{\partial t_{\rm gf}} = \nabla S[U(t_{\rm gf})] \cdot U(t_{\rm gf}) \, ,
\end{equation}
where $\nabla S[U(t_{\rm gf})]$ is the derivative of the action with respect to the gauge field, and the initial \(U(0)\) is the original lattice gauge field. 
It has been shown at a fixed \(\tau_{\rm gf}\), the original lattice cutoff can be removed, i.e., \(a\) taken to 0, keeping all physical matrix elements finite, provided only the parameters in the action and the wavefunction of the fermions are renormalized~\cite{Luscher:2010iy}. As opposed to the lattice regularization, this scheme has a number of advantages. In particular, as \(a\to0\), the symmetries of the continuum gauge theory, including Euclidean and chiral invariance~\cite{Luscher:2013cpa}, are restored. Furthermore, composite operators do not need additional regularization---their matrix elements are automatically finite~\cite{Luscher:2010iy}.  And finally, both the nonperturbative implementation of the gradient flow and the perturbation theory needed to connect to any other scheme are reasonably straightforward~\cite{Crosas:2023anw,Buhler:2023gsg}.

\section{Topological charge and Weinberg operators}
The relationship between the gradient flow (gf) scheme and \(\overline{\rm MS}\) renormalization scheme for gluonic CPV operators such as the topological charge \(G\cdot \tilde G\) and the Weinberg operator 
\(G \cdot \tilde G \cdot G\) can be described through the following matrix equation:
\begin{equation}
\begin{pmatrix} G\cdot\tilde G \\ G \cdot \tilde G \cdot G \\ \cdots\end{pmatrix}_{\overline{\rm MS}}
= \begin{pmatrix}1&0&0\\
                 Z_{M}\frac1{(\tau_{\rm gf}a)^2}&Z_{W}&O((\tau_{\rm gf}a)^2)\\
                  O(\frac1{(\tau_{\rm gf}a)^4})&O(\frac1{(\tau_{\rm gf}a)^2})&\cdots
                 \end{pmatrix}
\begin{pmatrix} G\cdot\tilde G \\ G \cdot \tilde G \cdot G \\ \cdots\end{pmatrix}_{\rm gf} \, ,
\end{equation}
where, $Z_{W}$ is the renormalization constant for the Weinberg operator and $Z_{M}$ represents the renormalization constant for the mixed operator~\cite{Crosas:2023anw}. 
For regularization schemes like the gradient flow that keep the topological charge integral, the density \(G\cdot\tilde G \) is not renormalized.  In this preliminary work, we only present results in the gradient-flow scheme without conversion to \(\overline{\rm MS}\).

In this work, we will examine the topological, mixed, and Weinberg susceptibilities. These are defined as the second derivative of the renormalized effective action with respect to the corresponding renormalized potentials, i.e., \(\Theta\) and \(w\), the coefficients of the topological and Weinberg terms added to the action. These quantities are, therefore, automatically renormalized. In practice, we calculate them as the variances and covariances of the corresponding charges: the topological charge and the volume integral of the Weinberg operator. Expressed in this way, they are the product of two volume integrals of densities, and include contribution from the contact terms between the two densities. Though these contact terms diverge, and, in a general scheme, would need extra regularization when treated as a density, their volume integral that contributes to the susceptibility is finite.  In summary, once the individual topological charge density and the Weinberg operator are renormalized, calculating susceptibilities do not need any further normalization.

\section{Susceptibilities: Topological, Weinberg and Mixed}\
Susceptibilities play a crucial role in understanding the vacuum structure of QCD by quantifying the system's response to external perturbations. 

The topological susceptibility $\chi_{Q}$ measures fluctuations in the topological charge $Q$, defined as $Q = \int d^4x G \cdot \tilde G$, where $G \cdot \tilde G$ is the topological charge density. It is expressed as:
\begin{equation}
\chi_{Q} = \langle Q^2 \rangle - \langle Q\rangle^2 = {\partial^2 \ln Z_\Theta}/{\partial \Theta^2} \, ,
\end{equation}
where $Z_\Theta$ is the partition function, and $\Theta$ is the CP-violating coefficient of the topological charge in the action. The expectation value of the topological charge, \(\langle Q\rangle\), being CP-violating vanishes in the QCD vacuum, and, hence we have \(\chi_Q = \langle Q^2\rangle\). The Weinberg susceptibility is similarly defined as $\chi_{W}= [\int d^4x (G \cdot \tilde G) \cdot G]^2$ and measures the response of the QCD vaccuum as the Weinberg operator is turned on. The mixed susceptibility $\chi_{M} = [\int d^4x (G \cdot \tilde G)][\int d^4x (G \cdot \tilde G \cdot G)]$ quantifies the interplay between topological and Weinberg charges. 

\section{Lattice calculations, fits to susceptibilities, and results for the Peccei-Quinn term}
\subsection{Lattice setup and ensemble parameters}
We employ tadpole-improved~\cite{Lepage:1992xa} clover fermions~\cite{Sheikholeslami:1985ij} for both both valence and sea quarks with the lattice parameters of the 11 ensembles given in~\cref{tab:latt}.  This action removes all tree-level \(O(a)\) errors as well as the anomalously large \(O(\alpha_s a)\) contributions arising from the lattice action. The remaining discretization errors are therefore \(O(a^2)\) and \(O(\alpha_s a)\) and with \(O(1)\) coefficients. The range of lattice spacings and quark masses ensures robust control over systematic uncertainties, enabling reliable extrapolations to the physical continuum and infinite-volume limits.  
\begin{table}[ht]
\centering
\small 
\setlength{\tabcolsep}{4pt} 
\renewcommand{\arraystretch}{0.5} 
\begin{tabular}{lcccc}
\toprule
\textbf{Ensemble name} & \(a\ \mathrm{(fm)}\) & \(M_\pi\ \mathrm{(MeV)}\) & \(M_K\ \mathrm{(MeV)}\) & \(L^3 \times T\) \\ 
\midrule
C13 & 0.127(2) & 285(5) & 476(5) & \(32^3 \times 96\) \\
D5L & 0.094(1) & 268(3) & 512(5) & \(48^3 \times 128\) \\
D220 & 0.094(1) & 214(3) & 543(6) & \(48^3 \times 128\) \\
D6 & 0.0914(9) & 175(2) & 492(5) & \(48^3 \times 96\) \\
D7 & 0.091(1) & 170(2) & 491(5) & \(64^3 \times 128\) \\
E5 & 0.0728(8) & 272(3) & 575(6) & \(48^3 \times 128\) \\
E6 & 0.0707(8) & 223(3) & 539(6) & \(64^3 \times 192\) \\
E7 & 0.0706(7) & 167(2) & 538(6) & \(72^3 \times 192\) \\
E9 & 0.0700(7) & 128(2) & 521(5) & \(96^3 \times 192\) \\
F5 & 0.056(1) & 280(5) & 526(6) & \(64^3 \times 192\) \\
F6 & 0.0555(6) & 216(2) & 527(5) & \(72^3 \times 192\) \\
\bottomrule
\end{tabular}
\caption{Summary of lattice ensembles used in this work.  All measured numbers are preliminary~\cite{ourspectrum}.}
\label{tab:latt}
\end{table}

\subsection{Fitting methodology for lattice data}
The dependence of observables for \(\tau_{\rm gf}\ll \Lambda_{\rm QCD}^{-1}\) is amenable to perturbation theory. On the other hand, they suffer from artifacts when \(a \lesssim\tau_{\rm gf}\) (or when \(\tau_{\rm gf} \gtrsim La\))~\cite{Luscher:2010iy,Bhattacharya:2021lol}.  As a result, only a small range provides a reliable guide to the flow-time behavior---all we can assert is they are expected to be smooth functions of this variable.  For this reason, in this work we have chosen to model the \(\tau_{\rm gf}\) dependence of all quantities by cubic splines~\cite{spline1967}.

For modeling the chiral and lattice spacing dependence, we have, however, chosen low order polynomials---in particular the data are fit either with a three, 
\begin{equation}
\chi = s_0(t_{\rm gf}) + s_1(t_{\rm gf}) a^p + s_2(t_{\rm gf}) M_\pi^2 \, ,
\label{e:3d}
\end{equation}
or a four-parameter form:
\begin{equation}
\chi = s_0(t_{\rm gf}) + s_1(t_{\rm gf}) a^p + s_2(t_{\rm gf}) M_\pi^2 + s_3(t_{\rm gf}) a^p M_\pi^2 \, ,
\label{e:4d}
\end{equation}
with the power \(p=1\) or 2. The motivation for testing the \(p=1\) form is that over this range of lattice spacings and masses, the strong coupling constant \(\alpha_s\) is almost a constant, and the expected \(\alpha_s a\) dependence can be treated as an \(a\) dependence.  On the other hand, if these effects are small, the strong \(O(a^2)\) effects may dominate the behavior.  
As mentioned before, the coefficients in each case are modeled as a cubic spline, with the number and positions of knots in the splines optimized using the Akaike Information Criterion (AIC)~\cite{1100705}, balancing model complexity with the quality of fit to prevent overfitting~\cite{1311138}. Contributions from $M^2_K$ being heavier than physical and finite volume effects are assumed negligible in this preliminary analysis. 
The fits exclude the ensembles C13, D220, D5L, and D6 to avoid coarse lattice spacing, heavy pion masses or small volumes (see~\cref{tab:latt}).  
A summary of the the fit quality assessed through both AIC and \(\chi^2\) per degree of freedom, is provided in~\cref{tab:fit_results}.
\begin{table}[t]
    \centering
    \begin{tabular}{|c|c|c|c|c|c|c|}
        \hline
        \multirow{2}{*}{Fit Type} & \multicolumn{2}{c|}{$\chi_Q$ (Topological)} & \multicolumn{2}{c|}{$\chi_W$ (Weinberg)} & \multicolumn{2}{c|}{$\chi_M$ (Mixed)} \\
        \cline{2-7}
        & AIC & $\chi^2/\text{dof}$ & AIC & $\chi^2/\text{dof}$ & AIC & $\chi^2/\text{dof}$ \\
        \hline
        3-parameter Fit with \(a\) (Eq.~\ref{e:3d}) & 156.654 & 2.493 & 91.008 & 1.067 & 135.460 & 2.029 \\
        3-parameter Fit with $a^2$ (Eq.~\ref{e:3d}) & 142.890 & 2.189 & 86.288 & 0.977 & 130.329 & 1.922 \\
        4-parameter Fit with \(a\) (Eq.~\ref{e:4d}) & 106.234 & 1.382 & 86.081 & 0.886 & 129.756 & 1.937 \\
        4-parameter Fit with $a^2$ (Eq.~\ref{e:4d}) & 115.116 & 1.572 & 86.365 & 0.871 & 129.334 & 1.867 \\
        \hline
    \end{tabular}
    \caption{Summary of fit qualities for the susceptibilities $\chi_Q$, $\chi_W$, and $\chi_M$.
    }
    \label{tab:fit_results}
\end{table}

\subsection{Fits to the susceptibilities}
\begin{figure}[ht]
    \centering
    \includegraphics[width=0.58\textwidth]{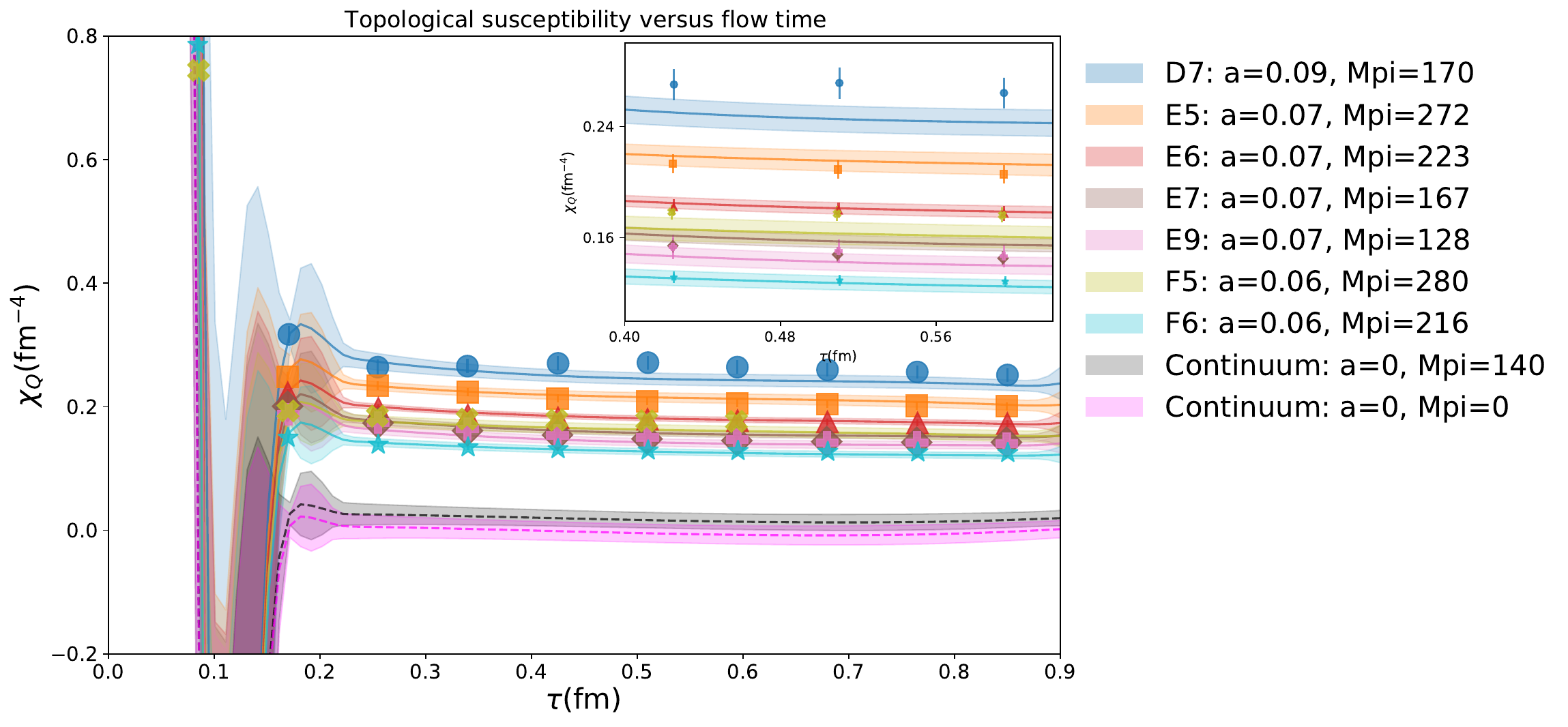}
    \includegraphics[width=0.41\textwidth,viewport=0 0 750 505,clip]{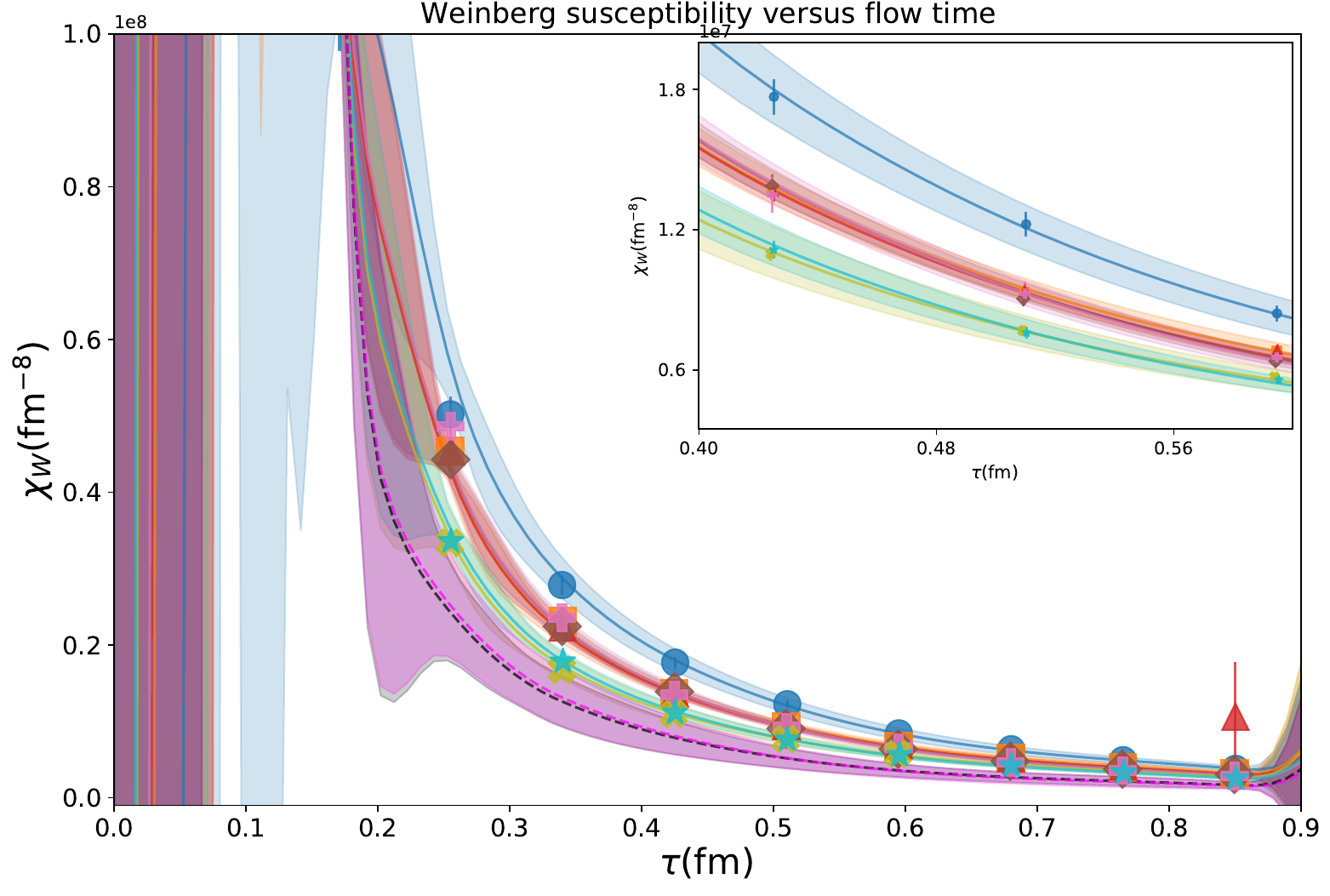}
    \caption{Topological (left) and Weinberg (right) susceptibility as a function of flow time for various lattice ensembles, and their chiral-continuum extrapolation. The data points represent lattice results for different ensembles, while the solid curves and error bars, obtained from fits, indicate the mean values with error bands corresponding to 1\(\sigma\) deviations. The overlapping magenta (chiral limit) and gray (physical quark masses) curves give the results after extrapolation in the quark masses and to the continuum limit.  }
    \label{fig:topological_charge}
    \label{fig:weinberg_susceptibility}
\end{figure}
Our preliminary results for the topological and Weinberg susceptibilities using the fit ansatz with $a^2$ dependence in Eq.~(\ref{e:3d}) are shown in Fig.~\ref{fig:weinberg_susceptibility}. The mixed susceptibility is shown in the left panel of Fig.~\ref{fig:PQ}.
Many fit forms were investigated---the choice of the results shown in Figs.~\ref{fig:weinberg_susceptibility}  and~\ref{fig:PQ} is motivated by the fit quality and the theoretical expectations that the topological susceptibility vanishes in the chiral- continuum limit~\cite{Bhattacharya:2021lol}. 
From these fits, we estimate the topological susceptibility to be approximately \(\rm (71 \, MeV)^4\) at the physical quark masses, in agreement with prior determinations~\cite{Bhattacharya:2023xov}. 

\subsection{Result for the Peccei-Quinn term \texorpdfstring{$\Theta_{\rm induced}$}{induced Theta}}
\begin{figure}[ht]
    \centering
    \includegraphics[width=0.6\textwidth]{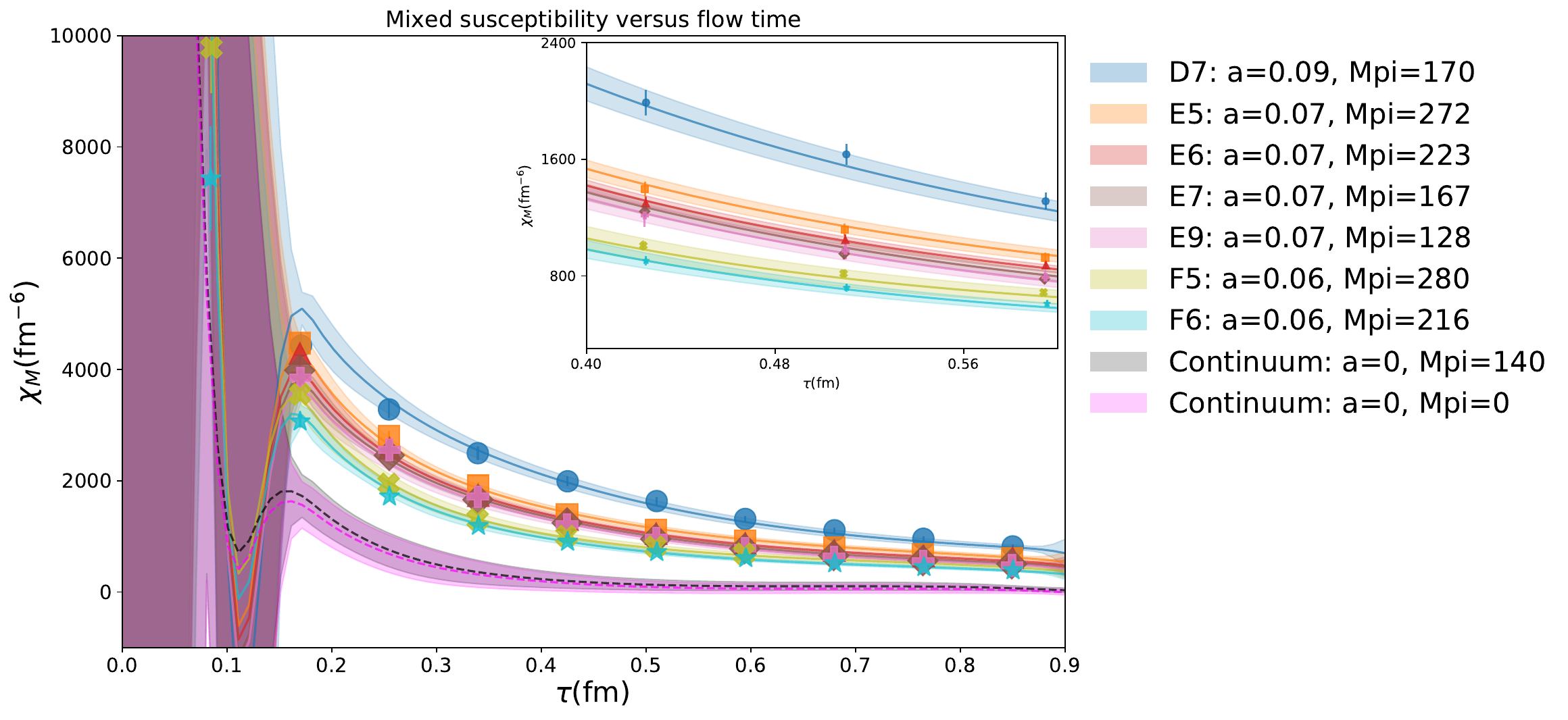}
    \includegraphics[width=0.39\textwidth]{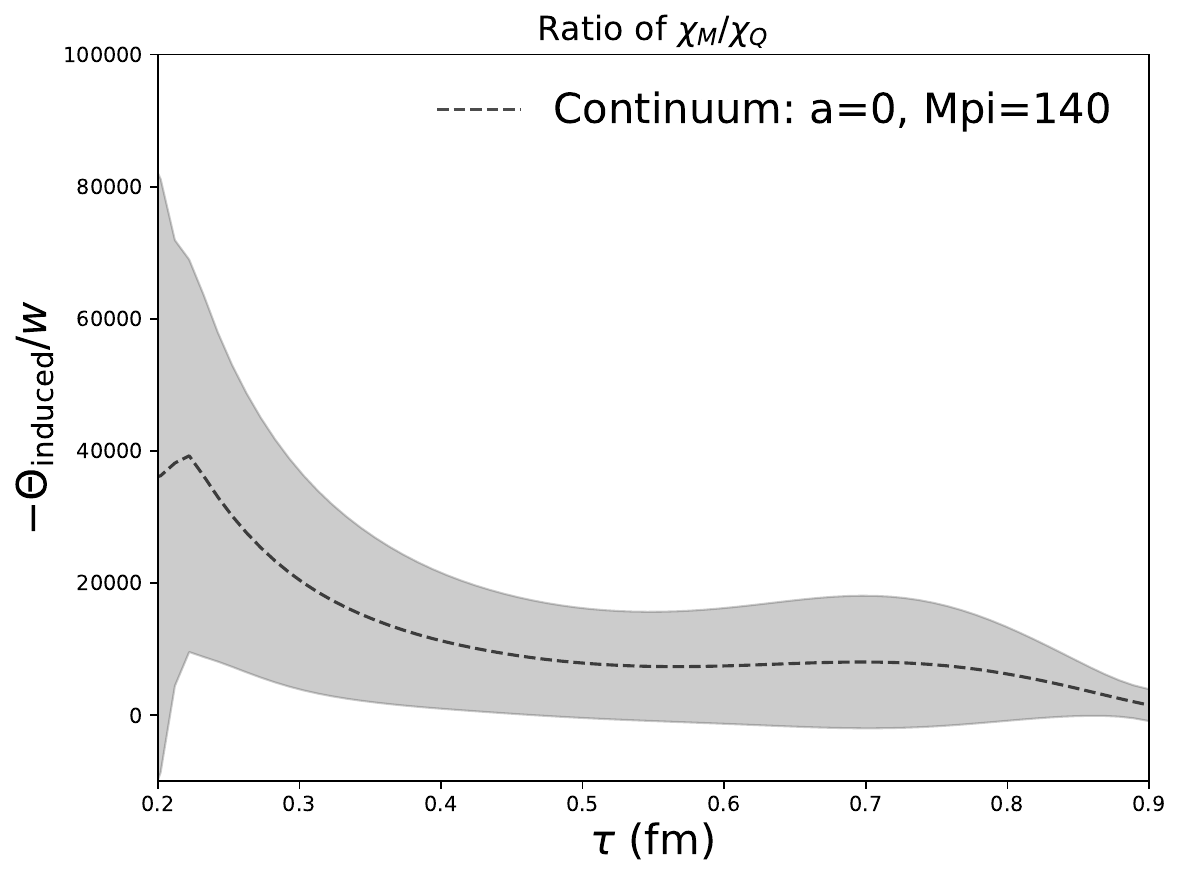}
    \caption{The chiral-continuum extrapolation of the mixed susceptibility is shown on the left using the same fit forms and ensembles as in \cref{fig:topological_charge}.  Dividing the extrapolated result by the topological susceptibility shown in \cref{fig:topological_charge}, one obtains the right panel showing the determination of $-\Theta_{\rm induced}/w$ relevant to PQ theories.}
    \label{fig:PQ}
\end{figure}
In theories using the Peccei-Quinn (PQ) mechanism, \(\Theta\), promoted to a field, dynamically relaxes to minimize its effective potential~\cite{Peccei:1977hh}.  If the Weinberg term is the only other source of CP violation, then this minimum is at~\cite{Pospelov:2000bw} 
\begin{align}
    \Theta_{\rm induced} = - w ({\chi_{M}}/{\chi_{Q}}) \, .
\end{align}
This expression is ill-defined in the chiral-continuum limit since both the susceptibilities vanish. In this limit, the  topological charge density can be rotated away, so the value of \(\Theta_{\rm induced}\) does not affect any physical matrix elements.  Our goal is to calculate it for the physical quark masses in the continuum limit. Unfortunately, data in Fig.~\ref{fig:PQ} show that the statistical uncertainties in this preliminary study are still too large to obtain a useful estimate of \(\Theta_{\rm induced}\).

\section{Summary}
Preliminary results on the gradient flow analysis of the Weinberg three-gluon operator are presented. The flow-time dependence of the lattice results for the topological, Weinberg, and mixed susceptibilities are fit by a cubic spline, and extrapolated to the continuum limit for both the chiral and the physical quark masses theories. For the topological susceptibility, our analysis confirms it vanishes at the chiral-continuum point. For physical quark masses, we get an almost flow-time independent value of approximately \((71 \, \text{MeV})^4\), consistent with prior determinations. The Weinberg susceptibility has a strong flow-time dependence, even though in the chiral-continuum limit, this dependence is only expected to be logarithmic. The statistical uncertainties in our calculations are still too large to obtain an useful estimate of $\Theta_{\rm induced}/w$.\looseness-1

\acknowledgments
The conceptualization, development of methodology, resource provision, data curation, validation, administration and supervision was due to T.B. and R.G. The lattice software was developed and the investigation and formal analysis was done by B.Y., J.-S.Y., R.G. and S.P.. S.B. and T.B. developed the fitting strategy, software and carried out the analysis. T.B., V.C. and E.M. developed the theoretical tools, and along with S.B. and R.G., interpreted the results. S.B. created the illustrations and along with T.B., E.M., and R.G., prepared the draft. All authors agree with the contents of the final manuscript. 

This work was supported by the US DOE under
Contract No. DE-AC52-06NA25396, the LANL LDRD program, and in part by LLNL under Contract DE-AC52-07NA27344. S.P. acknowledges the support from the ASC COSMON project. The calculations used the CHROMA software suite~\cite{Edwards:2004sx}.
Simulations were carried out at (i) the NERSC supported by DOE under Contract No. DE-AC02-
05CH11231; (ii) the Oak Ridge Leadership Computing Facility, which is a DOE Office of Science
User Facility supported under Award No. DE-AC05-00OR22725 and was awarded through the
INCITE program project HEP133, (iii) the USQCD collaboration resources funded by DOE HEP,
and (iv) Institutional Computing at Los Alamos National Laboratory. 
\newpage
\appendix

\bibliographystyle{JHEP}
\bibliography{main}
\end{document}